# Antiferromagnonic Spin Transport from $Y_3Fe_5O_{12}$ into NiO

Hailong Wang[†], Chunhui Du[†], P. Chris Hammel[*] and Fengyuan Yang[*]

Department of Physics, The Ohio State University, Columbus, OH, 43210, USA

[†]These authors made equal contributions to this work

[*]Emails: hammel@physics.osu.edu; fyyang@physics.osu.edu

We observe highly efficient dynamic spin injection from $Y_3Fe_5O_{12}$ (YIG) into NiO, an antiferromagnetic (AF) insulator, via strong coupling, and robust spin propagation in NiO up to 100-nm thickness mediated by its AF spin correlations. Strikingly, the insertion of a thin NiO layer between YIG and Pt significantly enhances the spin currents driven into Pt, suggesting exceptionally high spin transfer efficiency at both YIG/NiO and NiO/Pt interfaces. This offers a powerful platform for studying AF spin pumping and AF dynamics as well as for exploration of spin manipulation in tailored structures comprising metallic and insulating ferromagnets, antiferromagnets and nonmagnetic materials.

PACS: 75.50.Ee, 75.70.Cn, 76.50.+g, 81.15.Cd

Spin transport in ferromagnetic (FM) and nonmagnetic materials (NM) has been extensively studied [1-5]. Pure spin currents driven from FMs to metals or semiconductors by ferromagnetic resonance (FMR) or thermal spin pumping have attracted especially intense interest [6-16]. Another important class of magnetic materials, antiferromagnets, are not expected to enable spin transport; thus, the possibility of spin transport by AF excitations remains largely unexplored. FMR spin pumping in FM/NM bilayers relies on transfer of angular momentum from the precessing FM magnetization to the conduction electrons in the NM to generate spin currents [6-12, 14, 15]. Insulating FMs are known [7] to support spin transport through magnon currents. Simultaneous spin and magnon accumulation at a NM/FM-insulator interface accompanied by the interconversion of spin current $J_s$ to magnon current $J_m$ has been predicted [17, 18]. AFs, both metallic and insulating, can also sustain propagating spin excitations, potentially allowing transport of spin current. Our recently demonstrated growth of high-quality YIG thin films which enable mV-level inverse spin Hall effect (ISHE) spin pumping signals [14, 15, 19] provides an effective platform for observation of spin transport in AFs.

We grow 20-nm YIG films on $Gd_3Ga_5O_{12}$ (111) substrates, followed by deposition of NiO and Pt layers using off-axis sputtering [14, 15, 19-24]. X-ray diffraction (XRD) scan of one of our YIG films in Fig. 1(a) shows clear Laue oscillations, demonstrating its high crystalline quality. Fig. 1(b) shows an FMR derivative absorption spectrum for one (YIG-1) of the 20-nm YIG films studied in this letter taken at radio-frequency (rf) $f = 9.65$ GHz and power $P_{rf} = 0.2$ mW with an in-plane magnetic field ($\textbf{\textit{H}}$), from which we obtain a peak-to-peak linewidth ($\Delta H$) of 8.5 Oe [25]. Atomic force microscopy (AFM) measurements of a bare YIG film and a YIG/NiO(20 nm) bilayer shown in Figs. 1(c) and 1(d) reveal root-mean-square (rms) roughness of 0.197 and 0.100 nm, respectively, demonstrating the smooth surfaces of both YIG and NiO.

Our spin pumping measurements are performed at room temperature on ~1 mm × 5 mm samples in an FMR cavity with a DC field applied along the short edge, as illustrated in Fig. 1(e). For YIG(20 nm)/Pt(5 nm) bilayers at YIG resonance, the dynamical coupling between the precessing YIG magnetization and the conduction electrons in Pt produces a pure spin current, $J_s$, into Pt, which is converted to a net charge current via the ISHE [8-10, 26], resulting in an ISHE voltage ($V_{ISHE}$) along the length of the samples. Figure 1(f) shows $V_{ISHE}$ vs. $H$ - $H_{res}$ spectra, where $H_{res}$ is the FMR resonance field, at $\theta_H$ = 90° and 270° (two in-plane fields), which exhibits an ISHE voltage of 3.04 mV, the highest value we have observed [14, 15].

In this letter, we focus on three series of YIG/Pt bilayers and YIG/NiO/Pt trilayers prepared from three 20-nm YIG films labeled YIG-1, YIG-2 and YIG-3 with FMR linewidths (bare YIG) of 8.5, 10.4 and 22.6 Oe, respectively. The $\theta$-$2\theta$ XRD scan of a 100-nm NiO film deposited on GGG (111) substrate shown in Figure 1(g) indicates that the NiO films are polycrystalline with a preferred orientation along <111>. The top panels in Figs. 2(a)-2(c) show $V_{ISHE}$ vs. $H$ - $H_{res}$ spectra for the three YIG/Pt bilayers at $P_{rf}$ = 200 mW, which give $V_{ISHE}$ = 3.04 mV, 604 μV, and 146 μV, respectively. The three YIG/Pt bilayers are selected to have a wide range of ISHE voltages due to the difference in YIG film/interface quality.

To characterize spin transport in AF insulators, we insert a layer of NiO, an AF with a bulk Néel temperature over 500 K, between YIG(20 nm) and Pt(5 nm) in all three YIG series. The insulating nature of the NiO films is confirmed by electrical measurements. The middle panels in Figs. 2a-2c shows $V_{ISHE}$ vs $H$ spectra for the three series of YIG/NiO(1 nm)/Pt trilayers. Strikingly, we observe a significant enhancement, relative to Pt directly on YIG, of the spin pumping signals in all three samples: $V_{ISHE}$ = 4.71 mV (from 3.04 mV), 1.20 mV (from 604 μV), and 1.03 mV (from 146 μV), relative increases of 1.55, 1.99, and 7.05 for the YIG-1, YIG-2, and YIG-3 samples, respectively. Since the blocking temperatures ($T_b$) of 1- or 2-nm NiO films should be below room temperature [27, 28] ($T_b$ is expected to exceed 300 K at ~5 nm NiO thickness [29]), this indicates that the root of this enhancement of spin pumping efficiency in YIG/NiO/Pt trilayers lies in the AF fluctuations of NiO [30].

The dependence of the spin current injected into Pt on the NiO thickness provides clues as to length scale, and hence the mechanism underlying spin pumping observed here. Figures 3(a)-3(c) show semi-log plots of the dependencies of $V_{ISHE}$ on $t_{NiO}$ for the three series of trilayers. We observe three important features. First, for thin NiO interlayers, $t_{NiO} = 1$ or 2 nm, the ISHE voltages *increase* with increasing $t_{NiO}$. After peaking, the spin pumping signals of the trilayers remain higher than the values of corresponding YIG/Pt bilayers for $t_{NiO}$ up to 5 nm for the YIG-1 and YIG-2 series and up to $t_{NiO} > 10$ nm for the YIG-3 series. This is in notable contrast to the suppression of $V_{ISHE}$ by more than two orders of magnitude when a 1-nm nonmagnetic insulator SrTiO$_3$ (STO) is inserted between YIG and Pt, as shown in Fig. 3(d) [15]. The enhanced ISHE voltages suggest that the overall spin conversion efficiency [5, 10, 12] of the entire YIG/NiO/Pt trilayer is higher than the YIG/Pt bilayer with direct contact [14, 19], indicating that the YIG/NiO and NiO/Pt interfaces are exceptionally efficient in transferring spins. At the YIG/NiO interface, a strong, short-range exchange interaction [31] couples the FM magnetization in YIG with the AF moments in NiO [29]. At YIG resonance, the precessing YIG magnetization excites the AF moments at the YIG/NiO interface. The enhancement at $t_{NiO} \leq 2$ nm suggests that a prominent role for AF spin fluctuations in the spin transfer process.

Second, following this initial enhancement, $V_{ISHE}$ decays exponentially in all three series of YIG/NiO($t_{NiO}$)/Pt trilayers, implying diffusive spin transport in the AF insulator NiO. This presumably proceeds by means of either magnons (excitations of ordered AF spins when $T_b$ is above measurement temperature $T_m$) or AF fluctuations (excitations of dynamic but AF correlated spins when $T_b$ is below $T_m$). Least-squares fits to $V_{ISHE} = V_{ISHE}(t_{NiO} = 1 \text{ nm})e^{-t_{NiO}/\lambda}$ in the range 1 nm $\leq t_{NiO} \leq$ 50 nm indicate diffusion lengths $\lambda$ = 8.8, 9.4, and 11 nm for the YIG-1, YIG-2, and YIG-3 series, respectively, as compared to $\lambda$ = 0.19 nm for the YIG/STO/Pt trilayers. The AF magnons or fluctuations in NiO carry the angular momentum across the NiO thickness to the NiO/Pt interface, where the angular momentum is transferred across the NiO/Pt interface to the conduction electrons of Pt, generating a spin current in Pt.

Lastly, the decay of $V_{ISHE}$ slows at $t_{NiO} >$ 50 nm relative to thinner NiO. The bottom panels in Figs. 2(a), 2(b) and 2(c) show the $V_{ISHE}$ vs. $H - H_{res}$ spectra for the three YIG/NiO(100 nm)/Pt trilayers which give $V_{ISHE}$ = 1.85, 0.61 and 0.51 μV, respectively. The insulating nature of YIG and NiO rules out anisotropic magnetoresistance or anomalous Hall effect [32]. Magnetic proximity effect in Pt is not expected given that Pt is on top of antiferromagnetic NiO [33]. The slow decay in thick NiO suggests longer decay length in ordered AF.

Meanwhile, the YIG/NiO exchange coupling also induces extra damping in YIG which broadens the FMR linewidth. Figures 3(e)-3(g) show the NiO thickness dependence of $\Delta H$ for the three series of YIG/NiO/Pt trilayers, all of which exhibit an initial decrease in $\Delta H$, followed by an increases and eventual saturation at large NiO thickness. This behavior can be understood as follows. For very thin NiO (e.g., 1 or 2 nm), $T_b$ is well below room temperature and the AF fluctuation induced extra damping on YIG is small. However, an insulator as thin as 1 nm can effectively decouple YIG and Pt [see Fig. 3(g)] and greatly reduce the spin pumping induced extra damping by Pt. As a result, $\Delta H$ decreases first in very thin NiO regime. As NiO thickness increases, AF correlation becomes more robust and YIG/NiO exchange coupling grows stronger, which leads to an increase in damping and $\Delta H$. This is in clear contrast with YIG/SrTiO$_3$/Pt trilayers in Fig. 3(g) in which $\Delta H$ monotonically decreases before reaching saturation.

Exchange coupling between a FM and an AF can potentially lead to exchange bias and enhanced coercivity ($H_c$) [31]. Figure 4(a) shows the room temperature in-plane magnetic hysteresis loops of a 20-nm YIG film and YIG/NiO($t_{NiO}$) bilayers with $t_{NiO} = 2, 5, 10, 20$, and 50 nm. The bare YIG film exhibits a square hysteresis loop with a very small $H_c$ of 0.36 Oe and a very sharp magnetic switching with most of the reversal completed within 0.1 Oe, implying exceptionally high magnetic uniformity. As $t_{NiO}$ increases, $H_c$ continuously rises and reaches 1.53 Oe for YIG/NiO(50 nm), as shown in the inset to Fig. 4(a). We do not observe clear exchange bias in YIG/NiO bilayer (no exchange bias has been reported in YIG-based structures).

To verify that the observed spin transport across NiO in YIG/NiO/Pt trilayers does not arise from spurious effects, we grow four different heterostructures on YIG-1 and measure their spin pumping signals as shown in Fig. 4(b). The first sample, YIG/NiO(5 nm)/Cu(10 nm)/NiO(5 nm)/Pt(5 nm), in which we insert a 10-nm Cu spacer in between two 5-nm NiO layers, exhibits $V_{ISHE} = 1.25$ μV. This value is three orders of magnitude smaller than the value of 1.42 mV for YIG-1/NiO(10 nm)/Pt [Fig. 3(a)], indicating that spin current can still propagate from YIG to Pt across the three spacers, but the combined spin conductance of the three-layer/four-interface system is much smaller than for the YIG/NiO/Pt trilayers. Replacing the 10-nm Cu with a 5-nm SiO$_x$ layer in control sample (2) eliminates any detectable spin pumping signal, demonstrating that Cu can conduct spin current while SiO$_x$ blocks spin flow. The third control sample, YIG/NiO(5 nm)/Cu(10 nm), shows no ISHE signal, confirming that the observed spin pumping signal for YIG/NiO/Cu/NiO/Pt indeed comes from the ISHE in Pt. Lastly, the YIG/Cu(10 nm)/NiO(10 nm)/Pt structure shows a small but clear ISHE signal of 0.20 μV, indicating that while YIG/Cu(10 nm)/NiO(10 nm)/Pt can still propagate spins, the spin transfer efficiency is not as high as that in YIG/NiO(5 nm)/Cu(10 nm)/NiO(5 nm)/Pt(5 nm).

Altogether, this suggests the following multiple-stage spin conversion in YIG/NiO(5 nm)/Cu(10 nm)/NiO(5 nm)/Pt(5 nm): 1) at the YIG/NiO interface, the precessing YIG magnetization injects angular momentum into the first NiO, producing AF excitations; 2) the AF excitations carry the angular momentum to the first NiO/Cu interface, where they are converted to a spin current in Cu carried by the conduction electrons; 3) the spin current in Cu propagates to the second Cu/NiO interface where it is converted back to AF excitations in the second NiO layer; 4) the AF excitations in the second NiO layer transfer the angular momentum to the interface with Pt, where they are converted to a spin current in Pt, resulting in an ISHE voltage.

Figure 4(c) shows the FMR derivative absorption spectra of a bare YIG-1, a YIG-1/NiO(20 nm) and a YIG-1/SiO$_x$(20 nym) bilayer measured at $f$ = 9.65 GHz, which reveal that a 20-nm NiO significantly broadens the linewidth while SiO$_x$ has essentially no effect on the YIG linewidth. This suggests that the AF ordering or AF fluctuations in NiO plays an important role in the damping of YIG. Figure 4(d) gives the frequency dependencies of $\Delta H$ for the three samples shown in Fig. 4(c), all of which exhibit a linear relationship with frequency. From least-squares fits to the data in Fig. 4(d), we obtain the Gilbert damping constant $\alpha$ = 5.9 × 10$^{-4}$, 5.9 × 10$^{-4}$, and 2.5 × 10$^{-3}$ for YIG-1, YIG-1/SiO$_x$, and YIG-1/NiO, respectively [34]. The 20-nm NiO clearly enhances the damping in YIG while the damping in YIG/SiO$_x$ is almost the same as in bare YIG. This indicates that the AF moments in NiO exchange couple to the YIG magnetization in a way similar to the exchange bias in FM/AF bilayers [31], which causes additional damping in the FM.

In summary, we report observation of spin transport in AF insulator NiO and significant enhancement of spin pumping signals with insertion of a thin NiO spacer between YIG and Pt.

The enhanced spin pumping indicates excellent spin conversion efficiency at the YIG/NiO and NiO/Pt interfaces as well as robust spin transport in NiO mediated by AF magnons or AF fluctuations. The magnitude of spin currents in NiO decreases exponentially with decay lengths of ~10 nm within 1 nm $\leq t_{NiO} \leq$ 50 nm. This result suggests a new path toward high-efficiency spin transport by engineering heterostructures involving AF, FM and NM materials.

We thank the anonymous referees for insightful comments and valuable suggestions. This work is supported by the Department of Energy through grant DE-FG02-03ER46054 (CHD and PCH) and by the Center for Emergent Materials at the Ohio State University, a NSF Materials Research Science and Engineering Center (DMR-0820414) (HLW and FYY). Partial support is provided by Lake Shore Cryogenics Inc. (CHD) and the NanoSystems Laboratory at the Ohio State University.

**Figure captions:**

**Figure 1.** (a) A $\theta$-$2\theta$ XRD scan of a 20-nm YIG epitaxial film on $Gd_3Ga_5O_{12}$ (111) substrate near the YIG (444) peak. (b) A room-temperature FMR derivative absorption spectrum of a 20-nm YIG film (YIG-1) with an in-plane DC magnetic field and microwave power $P_{rf} = 0.2$ mW; the linewidth is 8.5 Oe. AFM images of (c) a 20-nm bare YIG film and (d) a YIG/NiO(20 nm) bilayer over an area of 10 μm × 10 μm, which exhibit rms roughness of 0.197 and 0.100 nm, respectively. (c) Schematic of ISHE measurement on YIG/Pt bilayer and YIG/NiO/Pt trilayers. (d) $V_{ISHE}$ vs. $H - H_{res}$ spectra at $\theta_H = 90°$ and 270° (two opposite in-plane fields) at $P_{rf} = 200$ mW for a YIG-1/Pt(5 nm) bilayer. (e) $\theta$-$2\theta$ XRD scan of a 100-nm NiO film on GGG(111) substrate.

**Figure 2.** $V_{ISHE}$ vs. $H - H_{res}$ spectra of YIG(20 nm)/NiO($t_{NiO}$)/Pt(5 nm) heterostructures at $P_{rf} = 200$ mW using (a) YIG-1, (b) YIG-2, and (c) YIG-3 with differing characteristics. The top, middle, and bottom panels are for YIG/Pt bilayers, YIG/NiO(1 nm)/Pt, and YIG/NiO(100 nm)/Pt trilayers, respectively.

**Figure 3.** Semi-log plots of the NiO thickness dependencies of the ISHE voltages for YIG(20 nm)/NiO($t_{NiO}$)/Pt(5 nm) trilayers using (a) YIG-1, (b) YIG-2, and (c) YIG-3. Inset: $V_{ISHE}$ as a function of NiO thickness from 0 to 100 nm for the three series of samples, where the horizontal dashed lines mark the values of $V_{ISHE}$ for the YIG/Pt bilayers. (d) Semi-log plot of $V_{ISHE}$ as a function of the $SrTiO_3$ barrier thickness for YIG/$SrTiO_3$/Pt trilayers. FMR linewidths as a function of NiO thickness for (e) YIG-1, (f) YIG-2, and (g) YIG-3 based trilayers, and (h) as a function of $SrTiO_3$ thickness in YIG/$SrTiO_3$/Pt trilayers.

**Figure 4.** (a) Room temperature magnetic hysteresis loops of a single YIG(20 nm) film and YIG/NiO bilayers with NiO thicknesses of 2, 5, 10, 20, and 50 nm, which give coercivities of 0.36, 0.42, 0.67, 0.93, 1.31, and 1.53 Oe, respectively. The inset shows the NiO thickness

dependence of coercivity. (b) $V_{ISHE}$ vs. $H$ - $H_{res}$ spectra of four heterostructures, including: (1) YIG(20 nm)/NiO(5 nm)/Cu(10 nm)/NiO(5 nm)/Pt(5 nm) (black), (2) YIG/NiO(5 nm)/SiO$_x$(5 nm)/NiO(5 nm)/Pt(5 nm) (purple), (3) YIG/NiO(5 nm)/Cu(10 nm) (blue), and (4) YIG/Cu(10 nm)/NiO(10 nm)/Pt(5 nm) (red), taken at $P_{rf}$ = 200 mW. The spectra are offset for clarity. (c) FMR derivative absorption spectra taken at $f$ = 9.65 GHz and (d) frequency dependencies of FMR linewidth of a bare YIG-1 film, a YIG-1/NiO(20 nm) and a YIG-1/SiO$_x$(20 nm) bilayer.

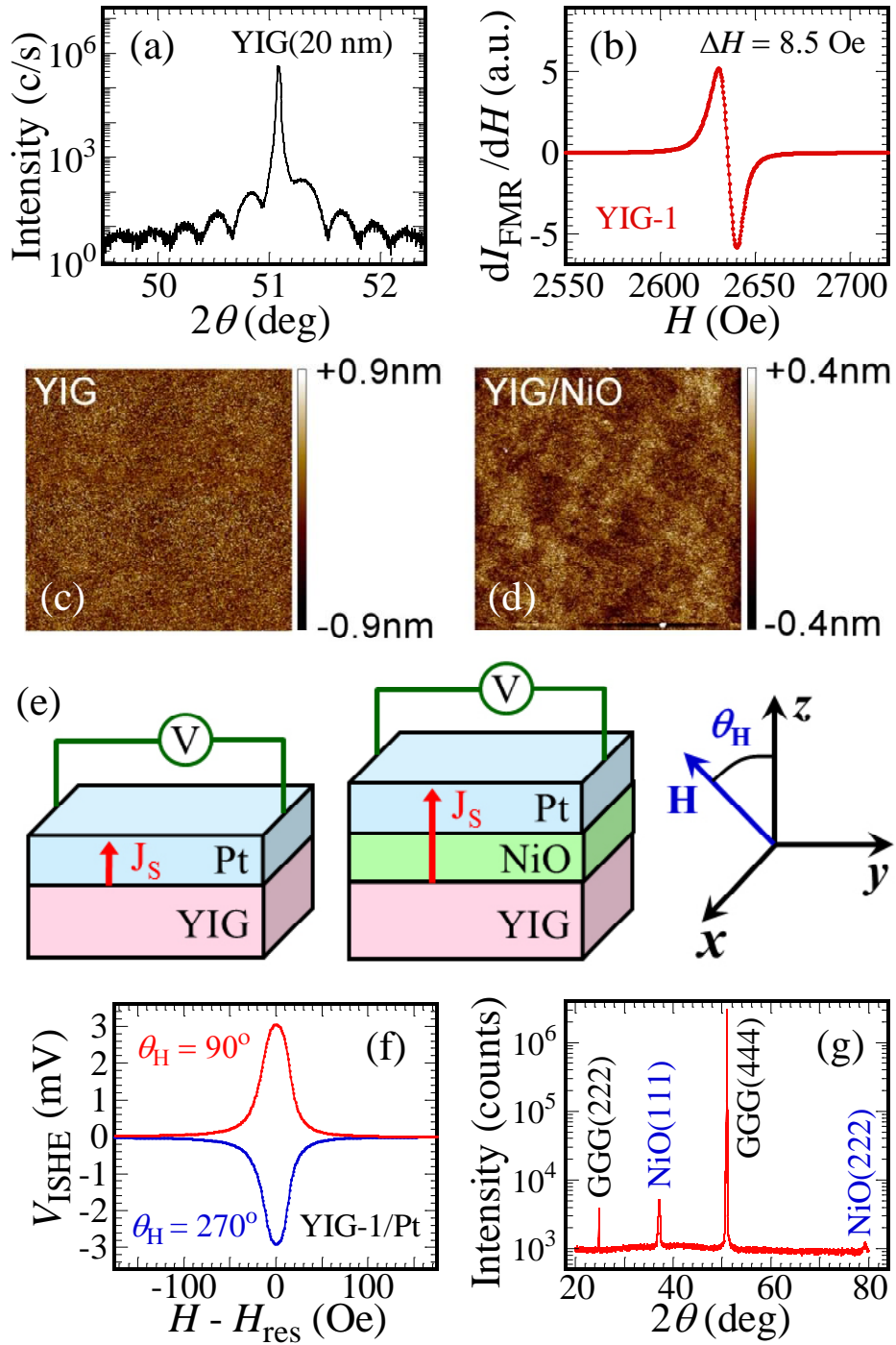

**Figure 1.**

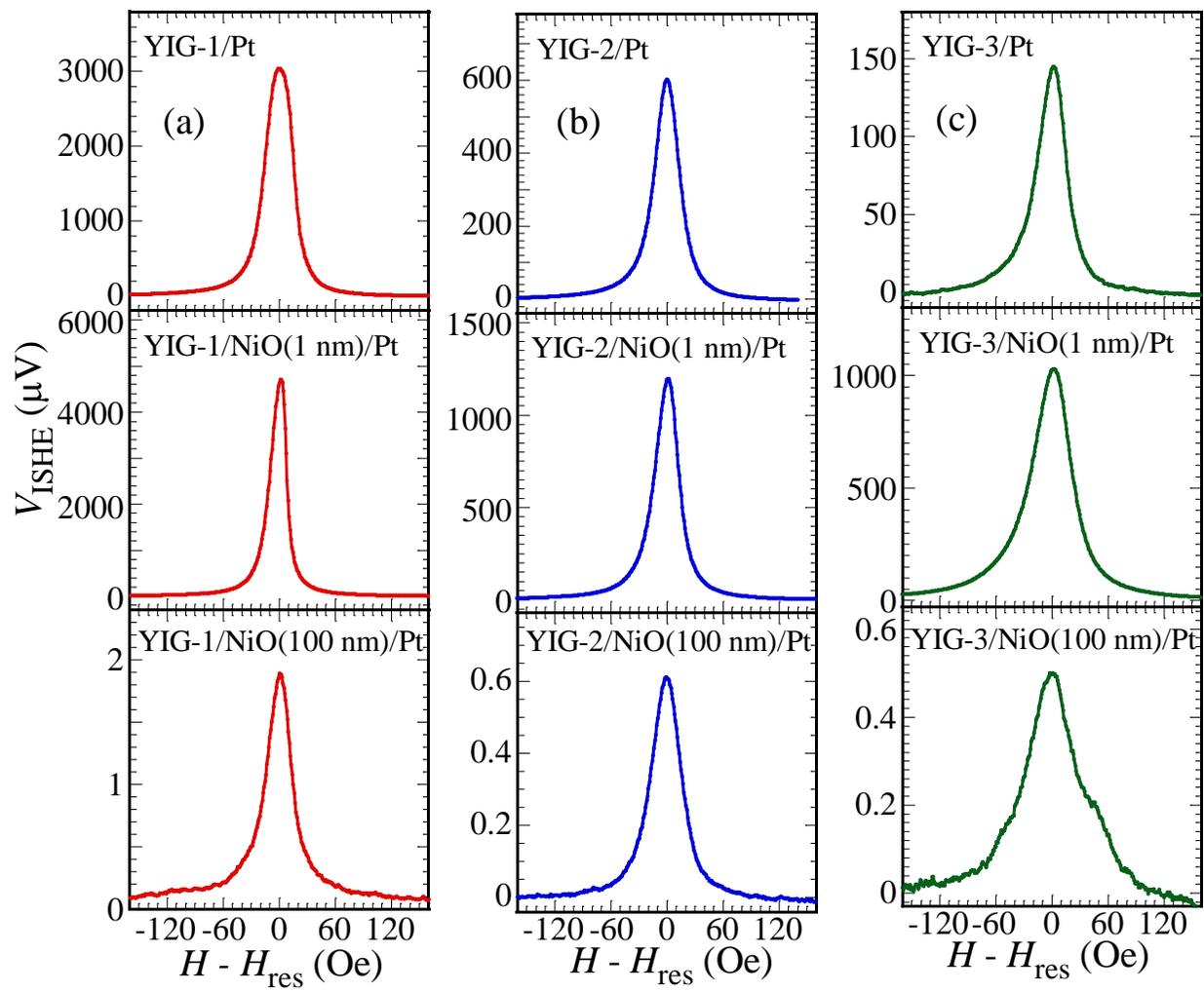

**Figure 2.**

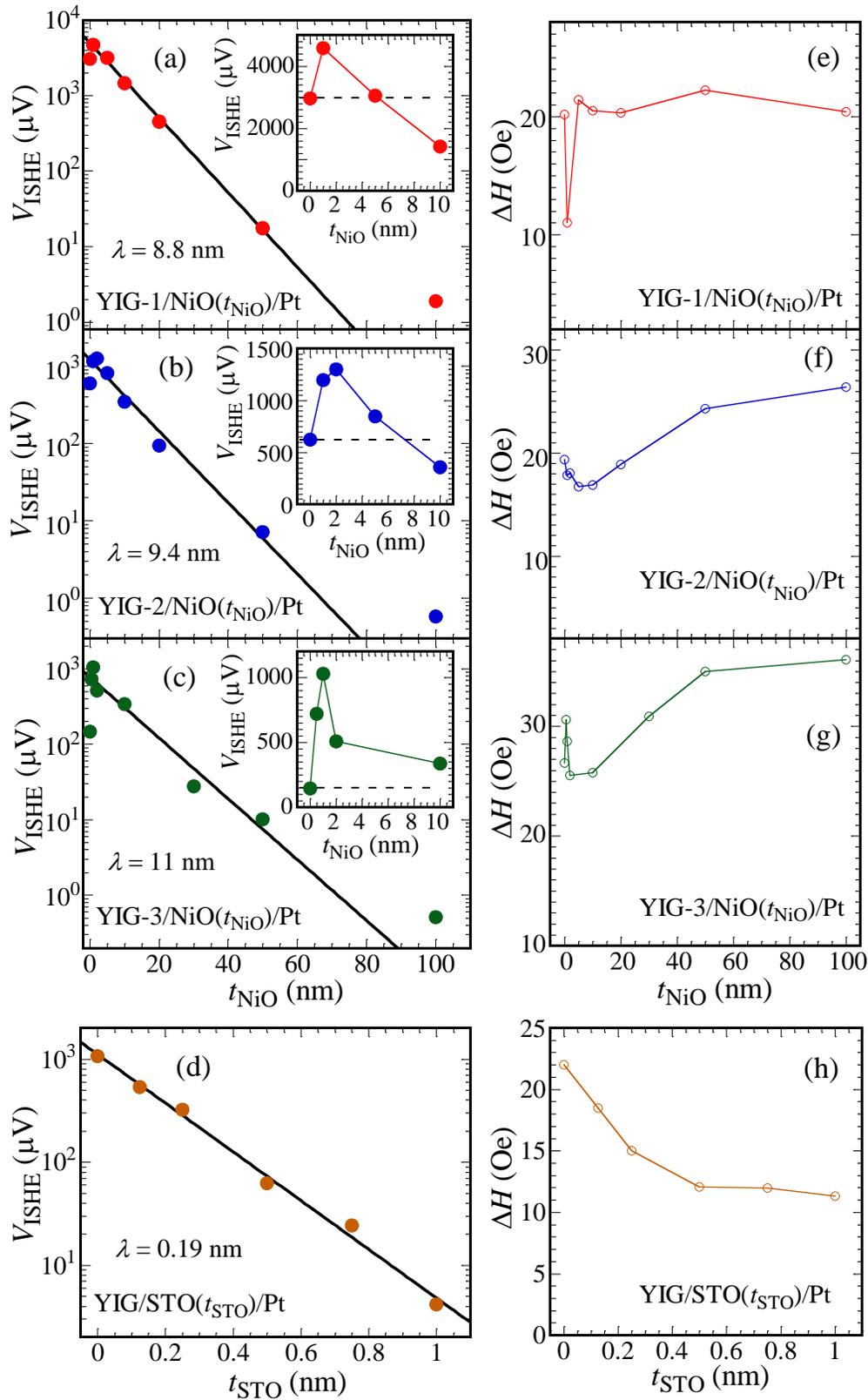

**Figure 3.**

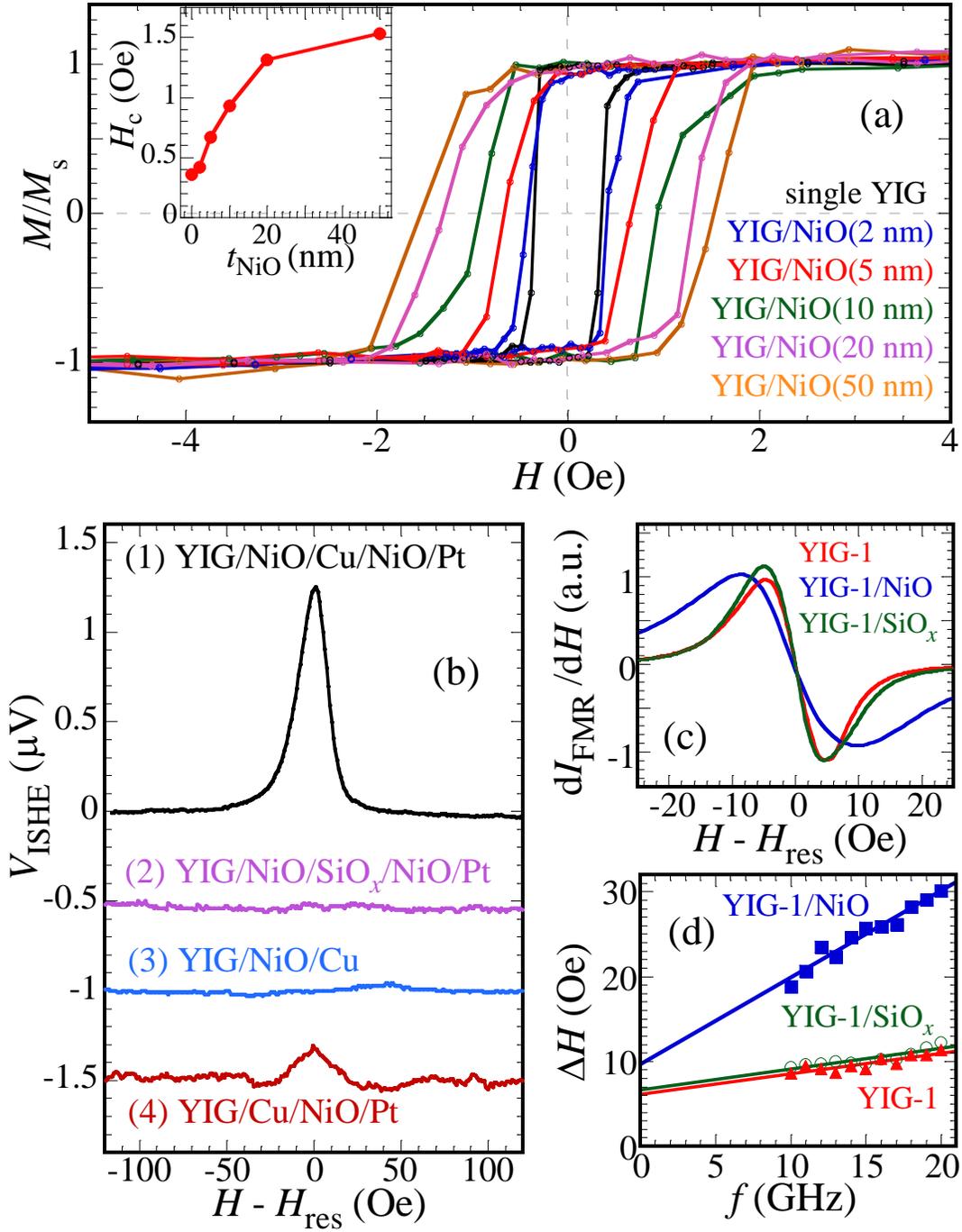

**Figure 4.**